\documentclass[a4paper,fleqn,usenatbib]{mnras}
\usepackage{graphicx}
\usepackage[normalem]{ulem}
\usepackage{hyperref}	
\hypersetup{colorlinks=true,linkcolor=blue,citecolor=blue,filecolor=blue,urlcolor=blue}
\usepackage{color}
\usepackage{upgreek}	
\usepackage{subcaption}
\usepackage{rotating}
\usepackage{amssymb}	
\newcommand{\be}{\begin{equation}}
\newcommand{\ee}{\end{equation}}
\newcommand{\bea}{\begin{eqnarray}}
\newcommand{\eea}{\end{eqnarray}}

\title
[Hot pairs in FR~II jets]
{On the significance of relativistically hot pairs in the jets of FR~II radio galaxies}

\author[Sikora, Nalewajko \& Madejski]{
Marek Sikora$^{1}$\thanks{E-mail: sikora@camk.edu.pl},
Krzysztof Nalewajko$^{1}$,
Greg M. Madejski$^{2}$
\\
$^{1}$Nicolaus Copernicus Astronomical Center, Polish Academy of Sciences, Bartycka 18, 00-716 Warsaw, Poland
\\
$^{2}$Kavli Institute for Particle Astrophysics and Cosmology, Stanford University, Stanford, CA 94305, USA
}

\begin{document}
\label{firstpage}
\pagerange{\pageref{firstpage}--\pageref{lastpage}}
\maketitle

\begin{abstract}
The energetic composition of radio lobes in the FR~II galaxies --- estimated by comparing their radio luminosities with the powers required to inflate cavities in the external medium --- seems to exclude the possibility of their energetic domination by protons.
Furthermore, if the jets were dominated by the kinetic energy of cold protons, it would be difficult to efficiently accelerate leptons in the jets' terminal shocks.
Assuming that the relative energy contents of leptons, protons and magnetic fields are preserved across the shocks, the above implies that the large-scale jets should also be energetically dominated by leptons: $P_{\rm e,j} \gtrsim P_{\rm p,j}$.
On the other hand, previous studies of small-scale jets in blazars and radio cores suggest a pair content (number of electrons and positrons per proton) of the order of $n_{\rm e}/n_{\rm p} \sim 20$.
Assuming further that the particle composition of jets does not evolve beyond the blazar scales, we show that this implies an average random Lorentz factor of leptons in large-scale jets of $\bar\gamma_{\rm e,j} \gtrsim 70(1+\chi_{\rm p})(20n_{\rm p}/n_{\rm e})$, and that the protons should be mildly relativistic with $\chi_{\rm p} \equiv (\epsilon_{\rm p} + p_{\rm p})/\rho_{\rm p} c^2 \lesssim 2$, $p_{\rm p}$ the pressure of protons, $\epsilon_{\rm p}$ the internal energy density of protons, and $\rho_{\rm p} c^2$ the rest-mass energy density of protons.
We derive the necessary conditions for loading the inner jets by electron-positron pairs and proton-electron plasma, and provide arguments that heating of leptons in jets is dominated by magnetic reconnection.
\end{abstract}


\section{Introduction}
\label{sec_intro}

Calorimetry of luminous radio lobes associated with some radio galaxies and
quasars indicates that they must be powered at the rates corresponding
with their AGN accretion powers \citep[e.g.,][]{1991Natur.349..138R,2007MNRAS.374L..10P,2011MNRAS.411.1909F,2017MNRAS.466.2294R}.
Such energy is transmitted from AGN to radio lobes by relativistic jets.
These jets have been observed in many spectral bands
(radio, infrared, optical, X-ray) on distance scales ranging from
sub-parsecs up to hundreds of kiloparsecs
\citep[see the review by][]{2019ARA&A..57..467B}.
Despite steady progress in the multi-band coverage, sensitivity and angular resolution of these observations,
the physical structure of jets, and its dependence on
the distance from the central black hole (BH), remain unclear.
The main reasons for this are:
(1) a rather weak dependence of the morphology of radio lobes on the matter 
content and magnetisation of the underlying jets \citep[e.g,][]{2010MNRAS.402....7M};
(2) a variety of dissipation mechanisms which can lead to similar
radiative properties of jets \citep[e.g.,][and refs. therein]{2019ARA&A..57..467B};
(3) an unknown efficiency of loading jets by mass \citep[e.g.,][]{2018ApJ...853...44O};
(4) a poor observational knowledge of the jets lateral structure \citep[e.g.,][]{2019BAAS...51c..59P}.

Relativistic speeds and powers of jets in luminous radio galaxies and
quasars associated with the FR~II radio sources \citep{1974MNRAS.167P..31F} seem to support the production of jets by the Blandford-Znajek mechanism \citep{1977MNRAS.179..433B} involving Magnetically Arrested Disks \citep[MAD;][]{2003PASJ...55L..69N,2009ApJ...704.1065P,2012MNRAS.423.3083M}.
In such a case,
the magnetic flux threading the BH is maximised, and the rate at which
the rotational energy of rapidly rotating BHs can be extracted and converted by
magnetic stresses to the kinetic energy of the outflows can reach or even
exceed the accretion power \citep{2011MNRAS.418L..79T}.
Such outflows are initially dominated by the
Poynting flux, but within $\sim 2-4$ distance decades about half of this flux
is theoretically predicted to be converted to the kinetic energy of cold
protons \citep{1994ApJ...426..269B,2010MNRAS.402..353L}.
One might try to verify this by studying blazars,
which emit most of their beamed radiation just at these distances \citep[e.g.,][]{2014ApJ...789..161N,2015MNRAS.449..431J}.

The observed spectral energy distributions (SED) of blazars can be in most cases reproduced by either the so-called `leptonic' models or the `(lepto)-hadronic' models \citep{2013ApJ...768...54B}.
The main difference is that, since the radiative efficiency of energetic protons is systematically lower than that of electrons/positrons \citep{2009ApJ...704...38S}, hadronic models require minimum jet powers larger typically by a factor $\sim 100$.
In many cases
(especially for the more luminous flat spectrum radio quasars, FSRQs, associated with the FR~II radio sources)
the hadronic jet power would exceed the Eddington luminosity corresponding to given BH mass by a factor up to tens/hundreds \citep{2015MNRAS.450L..21Z}, which would be in sharp conflict with estimates of the jet powers based on the radio-lobe calorimetry
\citep{1991Natur.349..138R,1999MNRAS.309.1017W,2013ApJ...769..129S}.
However, even for the strictly leptonic SED models,
and even in the limit of completely cold protons,
the contribution of protons to the total jet power
can be dominant, provided that $\bar\gamma_{\rm e,j} < (n_{\rm p}/n_{\rm e})(m_{\rm p}/m_{\rm e})$,
where $n_{\rm e} = n_{\rm e^+} + n_{\rm e^-}$.
In the case of no pairs ($n_{\rm p} = n_{\rm e}$), the powers of
blazar jets calculated by \cite{2014Natur.515..376G} are found to exceed
by a factor $\sim 20$ the jet powers estimated using the calorimetry of
radio lobes
\citep{2014ApJS..215....5K,2016Galax...4...12S,2017ApJ...840...46I,2017MNRAS.465.3506P,2018ApJ...861...97F}
and radio core shifts studies \citep{2017MNRAS.465.3506P}.

Little information on the proton content is available from the studies of radiative
spectra of the large-scale jets, or from the radio lobes.
Relativistic jets are considered as candidate sites for production of ultra-high-energy cosmic rays (UHECR) \citep[e.g.,][]{2014PhRvD..90b3007M,2018ApJ...854...54R}, and it has been suggested that synchrotron emission of ultra-relativistic protons may explain the extended X-ray emission from kpc-scale jets \citep[e.g.,][]{2002MNRAS.332..215A}.
However, as we argue below, the vast majority of protons in relativistic jets should be sub-relativistic, and as such they would yield no observational signatures that could be directly detected.

One might try to recover
information about the proton content by studying the rates at which
matter can entrain the jet at its base via interchange instabilities.
However, because the jets are formed as strongly electromagnetically dominated
outflows, the efficiency of the proton loading cannot be self-consistently
estimated using the currently available general relativistic MHD numerical simulations
\citep{2018ApJ...853...44O}.
Jets can also be entrained by protons on larger distances,
as a result of their interactions with the external medium \citep{2019MNRAS.490.2200C},
or intrinsically -- by winds of stars present within the jet volume
\citep{1994MNRAS.269..394K,2014MNRAS.441.1488P}.
While the dynamical effects of jet loading by stellar winds are likely to be
important in case of low-power FR~I radio galaxies (RGs), in the case of
powerful jets in FR~II RGs and quasars they are expected to be negligible
\citep{2019Galax...7...70P}.

We have a clearer picture for the problem of loading jets by electron-positron
pairs.
They can be created within the volume of the jet base by $\gamma$-rays 
emitted in high temperature accretion disk coronae
\citep{1996ApJ...458..514L,1999MNRAS.305..181B,1999ApJ...523L..21Y,2019ApJ...880...40I}.
As it will be shown in this paper, the rate of pair creation required
to provide the number flux of pairs needed to explain the blazar radiation
and the leptonic energy content of radio lobes is achievable for reasonable
parameters of accretion flows.
However, as we already pointed out before,
even for the pair-dominated jets, in the sense that the number density of pairs
largely exceeds the number density of protons,
one can still have the jet power dominated by the kinetic energy of cold protons.

But this seems to be challenged by studies of luminous FR~II radio sources
showing that the energy content of radio lobes is likely to be dominated by pairs
\citep{2012ApJ...751..101K,2016MNRAS.457.1124K,2017MNRAS.467.1586I,2018ApJ...855...71S,2018MNRAS.474.3361T,2018MNRAS.476.1614C}.
One could argue that the jet powers may still be dominated by the kinetic
energy of cold protons by postulating that the kinetic energy of protons
dissipated at the terminal shocks is roughly evenly redistributed between protons and leptons.
However, the results of particle-in-cell (PIC) numerical simulations
of relativistic shocks suggest that this would be possible only in the case of
parallel shocks (with magnetic field lines parallel to the shock normal)
\citep[see][and refs. therein]{2011ApJ...726...75S,2019MNRAS.485.5105C},
while terminal shocks associated with the hot spots in radio lobes are expected to be perpendicular.

This motivated us to investigate a scenario in which
the power of relativistic jets on large scales
is not dominated by the energy flux of cold protons,
but rather by the flux of relativistically hot pair plasma
\citep{2018ApJ...855...71S}.
In \S\ref{sec_content} we derive the parameters of
such jets using the observational data and theoretical constraints imposed by
the studies of radio lobes and blazars;
in \S\ref{sec_loading} we investigate the conditions
which may lead to the formation of jets with a large pair content,
and in \S\ref{sec_discussion} we discuss possible mechanisms
which on large scales may lead to the domination
of the jet power by the leptonic component.
The main results of our study are summarised
in \S\ref{sec_summary}.

\section{Physical structure of large-scale relativistic jets}
\label{sec_content}

Assuming that redistribution of energy between protons, leptons and
magnetic fields at the jet terminal shocks is negligible --
which is likely  to be the case when the jet power is dominated by the internal
energy flux -- the ratios of the jet power components,
$P_{\rm e,j} : P_{\rm p,j} : P_{\rm B,j}$ follow the ratios
of the lobe energy components, $E_{\rm e,l} : E_{\rm p,l} : E_{\rm B,l}$.
This implies that
$\kappa_{\rm j} \equiv P_{\rm e,j}/P_{\rm p,j} \simeq \kappa_{\rm l} \equiv E_{\rm e,l}/E_{\rm p,l}$
and
$\sigma_{\rm j} \equiv P_{\rm B,j}/(P_{\rm e,j} + P_{\rm p,j}) \simeq \sigma_{\rm l} \equiv E_{\rm B,l}/ (E_{\rm e,l} + E_{\rm p,l})$,
which allows us to derive constraints on the jet parameters
imposed by the knowledge of $\kappa_{\rm l}$ and $\sigma_{\rm l}$ from the observations of radio lobes.

\subsection{Mean lepton energy}
\label{sec_lepton_energy}

The powers of a relativistic jet (with bulk Lorentz factor $\Gamma_{\rm j} \gg 1$),
carried by relativistically hot leptons (with mean random Lorentz factor $\bar\gamma_{\rm e,j} \gg 1$, hence with the adiabatic index of $4/3$) and by relativistically warm protons, are equal to the relativistic enthalpy fluxes
\be
P_{\rm e,j} \simeq \frac{4}{3} \Gamma_{\rm j}^2\bar\gamma_{\rm e,j} n_{\rm e} m_{\rm e} c^3 A
\label{eq_Pej}
\ee
and
\be
P_{\rm p,j} = (1+\chi_{\rm p}) \Gamma_{\rm j} \dot{M}_{\rm p,j} c^2
= (1+\chi_{\rm p}) \Gamma_{\rm j}^2 n_{\rm p} m_{\rm p} c^3 A\,,
\label{eq_Ppj}
\ee
respectively, where
\be
\chi_{\rm p}
= \frac{P_{\rm p,j}}{\Gamma_{\rm j} \dot{M}_{\rm p,j}c^2} - 1
= \frac{\epsilon_{\rm p} + p_{\rm p}}{\rho_{\rm p} c^2}\,,
\ee
$\dot{M}_{\rm p,j} \equiv \Gamma_{\rm j} n_{\rm p} m_{\rm p} c A$ is the mass outflow rate, $A$ is the jet cross section, $\epsilon_{\rm p}$ is the internal energy density of protons, $p_{\rm p}$ is the pressure of protons, and $\rho_{\rm p} c^2$ is the rest mass energy of protons.
This implies that
\be
\kappa_{\rm j} \equiv \frac{P_{\rm e,j}}{P_{\rm p,j}} =
\frac{4}{3} \frac{\bar\gamma_{\rm e,j}}{(1+\chi_{\rm p})} \frac{n_{\rm e} m_{\rm e}}{n_{\rm p} m_{\rm p}} \,,
\ee
which is constrained to
$\kappa_{\rm j} \simeq \kappa_{\rm l} \equiv E_{\rm e,l}/E_{\rm p,l} \gtrsim 1$
by observations of radio lobes that estimate their energy contents due to leptons $E_{\rm e,l}$ and protons $E_{\rm p,l}$
\citep[e.g.,][]{2005ApJ...622..797K,2005ApJ...626..733C,2017MNRAS.467.1586I,2018MNRAS.474.3361T,2018MNRAS.476.1614C}.
We adopt here a leptonic content of $n_{\rm e}/n_{\rm p} \sim 20$,
as suggested by studies of the cocoon dynamics
\citep[e.g.,][]{2012ApJ...751..101K},
as well as by studies based on comparison of energetics
of blazar and radio cores with energetics of radio lobes
\citep[e.g.,][]{2017MNRAS.465.3506P}.
With this, we obtain
\be
\bar\gamma_{\rm e,j} = 70\kappa_{\rm j} (1+\chi_{\rm p}) \left(\frac{20}{n_{\rm e}/n_{\rm p}}\right)\,.
\label{eq_gammae}
\ee

\subsection{The proton energy content}
\label{sec_proton_chi}

Without knowledge of $\chi_{\rm p}$,
Eq. (\ref{eq_gammae}) allows to derive only a lower limit
for $\bar\gamma_{\rm e,j}$, i.e., that for $\chi_{\rm p} = 0$.
However, the value of $\bar\gamma_{\rm e,j}$ can be estimated using radiation spectra of the hot spots,
which are emitted by the shocked jet plasma.
Studies of hot spots
indicate that
the minimum random Lorentz factor of their leptons
$\gamma_{\rm e,min,hs}$ is of the order of a few hundreds
\citep[e.g.,][]{2001A&A...373..881H,2005ApJ...630..721T,2007ApJ...662..213S,2009ApJ...695..707G,2016MNRAS.463.3143M,2018MNRAS.474.3361T}.
Considering the typical spectral indices $\alpha_{\rm hs} \simeq 0.75$ of synchrotron spectra $F({\nu}) \propto \nu^{-\alpha}$ produced in hot spots in the radio band,
we estimate the mean random Lorentz factors of the hot spot leptons to be
\be
\bar\gamma_{\rm e,hs} \simeq
\left(\frac{s_{\rm hs}-1}{s_{\rm hs}-2}\right) \gamma_{\rm e,min,hs}
\simeq 3\gamma_{\rm e,min,hs} \sim 10^3\,,
\ee
where $s_{\rm hs} = 2\alpha_{\rm hs}+1 \simeq 2.5$ is the power-law index of the lepton energy distribution $N({\gamma}) \propto \gamma^{-s}$.
Combined with our assumption that the energy contents of leptons, protons and magnetic fields are preserved
across the terminal shocks, we obtain
\be
\bar\gamma_{\rm e,j} \sim 100 \left(\frac{\bar\gamma_{\rm e,hs}}{10^3}\right) \left(\frac{\Gamma_{\rm j}}{10}\right)^{-1}\,.
\ee
With this, from Eq. (\ref{eq_gammae}) we find that
\be
\chi_{\rm p} \simeq \left(\frac{1.4}{\kappa_{\rm j}}\right) \left(\frac{\bar\gamma_{\rm e,j}}{100}\right) \left(\frac{n_{\rm e}}{20n_{\rm p}}\right) - 1\,. 
\label{eq_chi}
\ee
Noting that $\kappa_{\rm j} \simeq \kappa_{\rm l} \ge 1$,
that $n_{\rm e}/n_{\rm p} \gg 20$ would be inconsistent with studies of blazars and
would result in too strong Compton effect
\citep{2010MNRAS.409L..79G},
and that distribution of the jet Lorentz factor is typically
peaked between $\Gamma_{\rm j} \simeq 5$
and $\Gamma_{\rm j} \simeq 15$
\citep[see][and refs. therein]{2001ApJS..134..181J,2004ApJ...609..539K,2019ApJ...874...43L},
one may conclude from Eqs. (7) and (8) that $\chi_{\rm p} \lesssim 2$, i.e. that baryonic plasma is at most mildly relativistic.

\subsection{Poynting power}
\label{sec_poynting}

The amount of the Poynting flux, $P_{\rm B,j} \simeq \Gamma_{\rm j}^2(B^2/4\pi) c A$,
contributing on large scales to the total
jet power remains very uncertain
\citep[see, e.g.,][]{2005ApJ...625...72S,2010MNRAS.402....7M}.
Within the scenario considered here
it can be estimated by using constraints imposed on the magnetisation of the
leptonic plasma in radio lobes.
As radio and X-ray observations of radio lobes
indicate
\citep[e.g.,][]{1998ApJ...499..713T,2002ApJ...580L.111I,2005ApJ...626..733C,2005ApJ...632..781I,2005ApJ...622..797K,2009PASJ...61S.327T,2009ApJ...706..454I,2010MNRAS.404.2018H,2010ApJ...714...37Y,2020MNRAS.491.5740P},
$\sigma_{\rm e,l} \equiv E_{\rm B,l}/E_{\rm e,l}$ is
estimated to be within the range $0.01 - 1.0$,
and most often claimed to be of the order $\sim 0.3$.
Since
\be
\sigma_{\rm j} \equiv
\frac{P_{\rm B,j}}{P_{\rm e,j} + P_{\rm p,j}} \simeq \sigma_{\rm l} \equiv
\frac{E_{\rm B,l}}{E_{\rm e,l} + E_{\rm p,l}} =
\left(\frac{\kappa_{\rm l}}{1+\kappa_{\rm l}}\right) \sigma_{\rm e,l}\,,
\ee
for $\kappa_{\rm l} \sim 1$ the Poynting power of large-scale jets cannot dominate over their kinetic power.

\subsection{The baryonic and leptonic power components}
  
Combining the estimates of the total jet power
($P_{\rm j} =  P_{\rm e,j} + P_{\rm p,j} + P_{\rm B,j}$)
based on the calorimetry of the radio lobes
(see \S\ref{sec_intro})
with the constraints on
the parameters $\kappa_{\rm j}$ and $\sigma_{\rm j}$
(see \S\ref{sec_lepton_energy} and \S\ref{sec_poynting}),
one can
express the powers carried by leptons and protons
in the large scale jets as
\be
P_{\rm e,j} = \kappa_{\rm j} P_{\rm p,j} = \frac{\kappa_{\rm j}}{(1+\kappa_{\rm j})(1+\sigma_{\rm j})} P_{\rm j} \,,
\label{eq_Pej_Ppj_Pj}
\ee
where $\kappa_{\rm j} \simeq \kappa_{\rm l}$ and $\sigma_{\rm j} \simeq
\sigma_{\rm l}$.

\subsection{Particle number fluxes}
\label{sec_number_flux}

Using Eq. (\ref{eq_Pej}), one can estimate the leptonic number flux
\be
\label{eq_dotN_e_jet}
\dot{N}_{\rm e,j}
\equiv \Gamma_{\rm j} n_{\rm e} c A
= \left(\frac{\kappa_{\rm j}}{1+\kappa_{\rm j}}\right)
\frac{P_{\rm j}}{(4/3)(1+\sigma_{\rm j}) \Gamma_{\rm j} \bar\gamma_{\rm e,j} m_{\rm e} c^2} \,,
\ee
and the proton number flux
$\dot{N}_{\rm p,j} = (n_{\rm p}/n_{\rm e}) \dot{N}_{\rm e,j}$.

\section{Loading of jets by matter}
\label{sec_loading}

\subsection{Pair Loading} 
\label{sec_loading_pair}

For loading jets by electron-positron pairs in AGN characterised by high and moderate accretion rates,
we consider pair creation by $\gamma$-rays
emitted in the accretion disk coronae.
The injection rate of pairs due to the
photon-photon annihilation within the jet base can be estimated as
\be
\dot{N}_{\rm e(\gamma\gamma)} = 2 f_{\rm jb} \dot{N}_{\gamma(>{\rm MeV})} \, \tau_{\gamma\gamma}\,,
\ee
where $f_{\rm jb}$ is the fraction of the total number of pairs produced
by the AGN that are created within the volume occupied
by the jet base, 
$\dot N_{\gamma(>{\rm MeV})}$ is the emission rate of the $E > 1\;{\rm MeV}$ photons by the hot accretion flow or its corona,
and $\tau_{\gamma\gamma}$ is the absorption probability of these photons
due to the pair production process.
Using approximate formulae
\bea
\dot{N}_{\gamma(>{\rm MeV})} &\simeq& \frac{f_{\gamma(>{\rm MeV})} L_{\rm acc}}{m_{\rm e}c^2} \,,
\\
\tau_{\gamma\gamma} &\simeq& n_{\gamma(>{\rm MeV})} R_{\gamma} \sigma_{\gamma\gamma}
\simeq
\frac{f_{\gamma(>{\rm MeV})} L_{\rm acc} \sigma_{\gamma\gamma}}{4\pi R_{\gamma} m_{\rm e} c^3} \,,
\eea
where $L_{\rm acc}$ is the bolometric luminosity of the accretion flow,
$f_{\gamma(>{\rm MeV})}$ is the fraction of that luminosity contained in the $E > 1\;{\rm MeV}$ photons,
$R_{\gamma}$ is the approximate size of the region from which most of the $\gamma$-rays are emitted,
$n_{\gamma(>{\rm MeV})}$ is the mean number density of the $E > 1\;{\rm MeV}$ photons within $R_\gamma$,
and $\sigma_{\gamma\gamma}$ is the cross-section for the pair production process,
we obtain
\bea
\dot{N}_{\rm e(\gamma\gamma)}
&\simeq&
f_{\rm jb} f_{\gamma(>{\rm MeV})}^2
\frac{\sigma_{\gamma\gamma}}{2 \pi c R_{\gamma}}
\left( \frac{L_{\rm acc}}{m_{\rm e} c^2} \right)^2
\\\nonumber
&\simeq&
2 f_{\rm jb} f_{\gamma(>{\rm MeV})}^2 \, \frac{m_{\rm p}}{m_{\rm e}} \,
\frac{\sigma_{\gamma\gamma}}{\sigma_{\rm T}} \,
\frac{\lambda_{\rm Edd}}{(R_{\gamma}/R_{\rm g})} \,
\frac{L_{\rm acc}}{m_{\rm e} c^2} \,,
\eea
where $\lambda_{\rm Edd} \equiv L_{\rm acc} / L_{\rm Edd}$ is the Eddington ratio, with $L_{\rm Edd} = 4\pi m_{\rm p} c^3 R_{\rm g}/\sigma_{\rm T}$ the Eddington luminosity, with $R_{\rm g} = GM_{\rm BH}/c^2$ the gravitational radius for BH mass $M_{\rm BH}$, and $\sigma_{\rm T}$ the Thomson cross section.
Hence, in order to load the jet by leptons at a rate
$\dot{N}_{\rm e,j}$ (Eq. \ref{eq_dotN_e_jet}),
the fraction of the accretion luminosity emitted
at energies $E > 1\;{\rm MeV}$ should be
\bea
f_{\gamma(>{\rm MeV})} &\simeq& 0.032 \left[
\frac{(P_{\rm j}/L_{\rm acc}) \, (R_{\gamma}/10 R_{\rm g})}
{(f_{\rm jb}/0.1) \, (\lambda_{\rm Edd}/0.1) \, (\bar\gamma_{\rm e,hs}/10^3)}
\times\right.
\nonumber\\&&
\left.\times
\frac{\kappa_{\rm j}}{(1+\sigma_{\rm j})(1+\kappa_{\rm j})}
\right]^{\frac{1}{2}} \, .
\eea
where $\sigma_{\gamma\gamma} \simeq 0.2 \sigma_{\rm T}$
was adopted \citep{1987MNRAS.227..403S}.

\subsection{Proton loading}
\label{sec_loading_prot}

Combining Eqs. (\ref{eq_Ppj}) and (\ref{eq_Pej_Ppj_Pj}),
one can find that
the required mass loading rate of the jet by protons is
\be
\dot{M}_{\rm p,load} \simeq
\frac{P_{\rm j}}{(1+\kappa_{\rm j})(1+\sigma_{\rm j})(1+\chi_{\rm p})\Gamma_{\rm j}c^2} \,.
\ee
where $\chi_{\rm p}$ is given by Eq. (\ref{eq_chi}).
Comparing this with the mass accretion rate $\dot{M}_{\rm acc} = L_{\rm acc}/(\epsilon_{\rm acc}c^2)$, with
$\epsilon_{\rm acc} \sim 0.1$
the radiative efficiency of the accretion flow, we estimate the proton loading efficiency as
\be
\label{dotM_p_load_eff}
\frac{\dot{M}_{\rm p,load}}{\dot{M}_{\rm acc}}
\simeq 10^{-2}
\frac{(\epsilon_{\rm acc}/0.1)}{(\Gamma_{\rm j}/10)(1+\kappa_{\rm j})(1+\sigma_{\rm j})(1+\chi_{\rm p})}
\, \left(\frac{P_{\rm j}}{L_{\rm acc}}\right) \,.
\ee
We note that studies of the jet powers in FR~II RGs and quasars indicate that the distribution of $P_{\rm j}/L_{\rm acc}$ peaks around $\sim 0.1$
\citep[e.g.,][]{2013A&A...557L...7V,2017ApJ...840...46I,2020ApJ...900..125R}.

The mass loading of a jet is also contributed by the leptons,
but at a rate lower by factor
\be
\frac{\dot M_{\rm e,load}}{\dot M_{\rm p,load}} =
\frac{\dot N_{\rm e,j}  m_{\rm e}}{\dot N_{\rm p,j} m_{\rm p}} \simeq
0.01 \left(\frac{n_{\rm e}}{20n_{\rm p}}\right)
\ee
compared with the baryons.

\section{Discussion}
\label{sec_discussion}

The results presented in \S\ref{sec_content} and \S\ref{sec_loading}
were obtained under the assumption that
$P_{\rm j}$, $\dot{N}_{\rm e,j}$ and $\dot{N}_{\rm p,j}$
do not depend on distance from the jet base,
i.e., that the jet energy losses due to
radiation and work done against the environment are negligible,
and that loading of jets by matter
(both by protons and pairs) is dominated by processes near the BH.
The former can be justified,
since powerful jets in FR~II classical double
radio sources are rather straight
\citep[e.g.,][]{1984RvMP...56..255B},
and their radiation consumes at most half of the jet energy
\citep[e.g.,][]{2014Natur.515..376G}.
The latter is less certain,
especially in the case of protons,
which can be loaded not only near the jet base.
Instabilities between the jet and its environment (accretion wind or ISM) may develop over a broad range of distance scales
\citep{2019MNRAS.490.2200C}.
Unfortunately, the efficiency of such
loadings is difficult to estimate,
due to both limited capabilities of MHD simulations
and poor observational constraints on the external medium.
Jets can be also loaded by winds from stars present
within the jet volume,
however,
in the case of powerful jets such loading cannot
compete with the required rate $\dot{M}_{\rm p,load}$
derived in \S\ref{sec_loading_prot}
\citep{1994MNRAS.269..394K,2014MNRAS.441.1488P}.

There is also uncertainty regarding the location of jet loading by pairs.
The efficiency of pair production
by high energy photons
emitted by the hot disk coronae depends strongly on the fraction
of AGN radiation emitted above $1\;{\rm MeV}$,
which presently can be only estimated
by extrapolating spectra detected at lower energies
\citep{1996MNRAS.282..646G,2013MNRAS.431.2471B,2015MNRAS.447.1289P,2016MNRAS.461.3165B,2017ApJS..233...17R,2018MNRAS.480.2861G}.
We have to wait for the next generation of MeV missions
\citep{2019ApJ...880...40I}
to verify whether the required leptonic loading rate
$\dot{N}_{\rm e,j}$
can be achieved already at the jet base,
or does it also involve processes
operating at somewhat larger distances,
like those considered by \cite{1995ApJ...441...79B} ($\lesssim 10^2 R_{\rm g}$)
or \cite{2006MNRAS.372.1217S} ($\sim 10^4 R_{\rm g}$).

Obviously, even if the particle number fluxes,
and hence $n_{\rm e}/n_{\rm p}$,
are fixed beyond a certain distance from the BH,
all such parameters as $\sigma_{\rm j}$,
$\bar\gamma_{\rm e,j}$ and
$\kappa_{\rm j}$
are expected to depend on distance significantly\footnote{Note that the values of these parameters
estimated in previous sections
are their final, `asymptotic' values.},
driven by evolution of the jet structure,
which is shaped by the external plasma profile.
At the smallest scales jets are presumably confined
by the MHD winds from the accretion disk,
and take a parabolic shape
\citep[e.g.,][]{2005MNRAS.357..918B,2006MNRAS.367..375B,2008ApJ...679..990Z,2010MNRAS.402..353L}.
These inner jets are strongly dominated by the Poynting flux,
while the power carried by leptons is strongly suppressed at those scales
due to extremely efficient
radiative cooling by synchrotron and IC mechanisms.
However, as theoretical models of relativistic MHD
jets predict
\citep[e.g.,][]{2009ApJ...698.1570L},
and observations of fast blazar variabilities support
\citep[e.g.,][]{2014ApJ...796L...5N},
already at distances smaller than $10^4 R_{\rm g}$
a significant fraction of the magnetic energy
is converted to the kinetic and internal energies
of the matter.
Such a conversion proceeds efficiently
up to the distance where
$\sigma_{\rm j}$ drops to about unity.
But as studies of blazar spectra indicate
\citep{2015MNRAS.448.1060G},
even at distances $(10^4 - 10^5) R_{\rm g}$,
the average lepton energy
$\bar\gamma_{\rm e,j}$ is at most a few tens,
i.e., too small for getting $P_{\rm e,j} \gtrsim P_{\rm p,j}$ even for $\chi_{\rm p} = 0$.
This implies that heating of the leptonic plasma must
continue beyond the `blazar zone'.

At the distance scale of hundreds of parsecs,
i.e., within the cores of their host galaxies,
jet confinement is provided by the interstellar medium (ISM).
In that case, if the external pressure $p_{\rm ext}$
decreases with distance $r$ faster than
$p_{\rm ext} \propto r^{-2}$, the jets become conical
\citep[e.g.,][]{1991MNRAS.250..581F,1994MNRAS.269..394K,2009ApJ...698.1570L,2017MNRAS.469.4957B}.
There, radiative energy losses of leptons are negligible
\citep{2013ApJ...779...68S,2014ApJ...789..161N,2015MNRAS.449..431J},
however, the dissipative processes must still operate
in order to protect the plasma against adiabatic cooling
in the laterally expanding jets
\citep[see][and refs. therein]{2015MNRAS.453.4070P,2019MNRAS.485.1210Z}.

Finally, outside the galaxies,
confinement is provided by the roughly uniform pressure
of the cocoon inflated by the shocked jet matter
backflowing sideways from the jet terminal shock,
and hence the jets at these scales
would become roughly cylindrical
\citep[e.g.,][]{2011ApJ...740..100B,2016MNRAS.461L..46T}.
In this `cylindrical zone' of a jet,
the intrinsic jet properties are expected
to be independent of distance
(no net heating by dissipative processes associated with pressure matching of expanding flows; no adiabatic cooling; negligible radiative losses),
and hence they should be settled in the previous zones.

Several dissipative processes have been proposed
to operate in relativistic jets:
(i) internal shocks forming between jet portions
moving at different velocities
\citep[e.g.,][]{2001MNRAS.325.1559S};
(ii) external (oblique/reconfinement) shocks which mediate
the pressure balance between jets and their environment
\citep[e.g.,][]{1994MNRAS.269..394K,2009ApJ...699.1274B,2009MNRAS.392.1205N};
and (iii) magnetic reconnection
driven by turbulence in magnetised plasma\footnote{In relativistic jets,
turbulence can be sustained by a variety of instabilities
developing in the presence of toroidal magnetic fields
\citep[e.g.,][]{1998ApJ...493..291B,2018PhRvL.121x5101A,2019ApJ...884...39B},
shear layers \citep[for recent review, see][]{2019Galax...7...78R},
or recollimation shock waves \citep[e.g.,][]{2018NatAs...2..167G},
but also due to interactions of the jet with `internal' obstacles
(dense molecular clouds and/or Wolf-Rayet stars with strong winds;
see, e.g., \citealt{2019Galax...7...70P}).}
\citep[see][and refs. therein]{2019ApJ...886..122C,2019MNRAS.tmp.2965S,2020MNRAS.493..603Z}.
The key issue is whether any of these dissipative processes can
make the power of large scale jets dominated by the enthalpy
flux of the leptonic plasma.
As PIC simulations indicate,
this cannot be achieved in shocks
\citep[see, e.g.,][]{2011ApJ...726...75S,2012ApJ...755...68S},
but under certain conditions
($\sigma_{\rm j} \gtrsim 1$ and $n_{\rm e}/n_{\rm p} \gg 1$)
can be accomplished in the scenarios involving energy dissipation
in magnetic reconnection sites \citep{2019ApJ...880...37P}.

\section{Summary}
\label{sec_summary}

According to the most popular model of relativistic jets,
they are launched dominated by the Poynting flux, and are gradually converted
to become dominated by the cold protons
\citep[for review, see][and refs. therein]{2019ARA&A..57..467B}.
The kinetic energy of cold protons would be further dissipated in the terminal shocks, and converted to relativistically hot protons and ultra-relativistic leptons.
Prior to the terminal shocks, the jet magnetic fields are expected to be dominated by the toroidal component,
and hence such shocks are predicted to be `perpendicular'.
Recent PIC simulations of perpendicular shocks indicate that
acceleration of electrons/positrons in such shocks is very inefficient,
and most of the energy of such cold jets would be converted to
the internal energy of the protons
\citep{2011ApJ...726...75S,2019MNRAS.485.5105C}.
This seems to be challenged by detailed studies of the radio lobes,
which indicate that at least half of their internal energy is contributed
by the relativistic pairs
\citep[][and refs. therein]{2018ApJ...855...71S}.
This problem can be overcome by assuming that in the large-scale jets the dominant portion of the energy flux is carried by relativistically hot leptons.

As we demonstrated in \S\ref{sec_lepton_energy},
such a picture of relativistic jets is supported by a combination
of observational data on radio lobes and blazars.
Their leptonic contents are $n_{\rm e}/n_{\rm p} \sim 20$,
and the leptons and protons are
characterised by mean energy
$\bar\gamma_{\rm e,j} \sim 100$
and parameter
$\chi_{\rm p} \equiv (\epsilon_{\rm p}+p_{\rm p})/\rho_{\rm p}c^2 \le 1$,
respectively.
While loading the jets by pairs is likely to be established already at the jet
base, where pairs can be created by high energy photons produced in
accretion disk corona (see \S\ref{sec_loading_pair}), it is not clear whether
loading by proton-electron plasma,
at the estimated rate of $\sim 1$\% of the accretion rate,
can be achieved also at the jet base -- by interchange instabilities
developed between the accretion flow and the electromagnetic outflow,
or are dominated by  processes working on larger scales  (see \S\ref{sec_loading_prot}).

In order for the large scale jets to be dominated
by leptons also in terms of energy flux,
it is required that dissipative mechanisms maintain leptons
at the average random Lorentz factor of
$\bar\gamma_{\rm e,j} \gtrsim
100(20 n_{\rm p}/n_{\rm e})$.
At distance scales larger than tens of parsecs,
the radiative cooling of even ultra-relativistic electrons is inefficient,
and in order to maintain the energy flux of leptons,
the required heating rate is determined mainly by the need to
compensate the adiabatic losses
\citep{2015MNRAS.453.4070P,2019MNRAS.485.1210Z}.
Such heating can be mediated by the oblique/reconfinement shocks,
which regulate the pressure balance between the jet and its environment,
and are predicted to stimulate the kink instabilities
followed by a variety of particle acceleration mechanisms
\citep[e.g.,][]{2016MNRAS.461L..46T,2017MNRAS.469.4957B,2018PhRvL.121x5101A,2019MNRAS.482.2107D}.
We particularly favour the magnetic reconnection,
which under certain conditions, contrary to the shocks,
allows to convert most of the dissipated energy to the leptonic plasma
\citep{2019ApJ...880...37P}.

\section*{Acknowledgements}
We thank the Reviewers for helpful suggestions.
We acknowledge financial support by the Polish National Science Centre grants 2016/21/B/ST9/01620 and 2015/18/E/ST9/00580.

\section*{Data availability}
There are no new data associated with this article.

\label{lastpage}
\end{document}